\documentclass[preprint,prl,twocolumn,10pt,a4paper,byrevtex]{revtex4}%
\usepackage{amsfonts}
\usepackage{amsmath}
\usepackage{amssymb}
\usepackage{graphicx}%
\providecommand{\U}[1]{\protect\rule{.1in}{.1in}}

\begin{document}
\preprint{ }
\title[Short title for running header]{Current-induced domain-wall motion in synthetic antiferromagnets}
\author{D. Herranz, R. Guerrero, R. Villar, F.G. Aliev}
\affiliation{Dpto F\'{\i}sica Materia Condensada, C-III, Universidad Aut\'{o}noma de
Madrid, Cantoblanco, Madrid, Spain}
\author{A.C. Swaving, R.A. Duine}
\affiliation{Institute for Theoretical Physics, Utrecht University, Leuvenlaan 4, 3584 CE,
Utrecht, The Netherlands}
\author{C. van Haesendonck}
\affiliation{LVSM, Katholieke Universiteit Leuven, Belgium}
\author{I.Vavra}
\affiliation{IEE SAS, D\'{u}bravsk\'{a} cesta 9, 84239 Bratislava, Slovakia}

\begin{abstract}
Domain-wall magnetoresistance and low-frequency noise have been studied in
epitaxial antiferromagnetically-coupled [Fe/Cr(001)]$_{10}$ multilayers and
ferromagnetic Co line structures as a function of DC current intensity. In
[Fe/Cr(001)]$_{10}$ multilayers a \textit{transition from excess to suppressed
domain-wall induced 1/f noise} above current densities of $j_{c}\sim
2\times10^{5}A/cm%
{{}^2}%
$ has been observed. In ferromagnetic Co line structures the domain wall
related noise remains qualitatively unchanged up to current densities
exceeding $10^{6}A/cm%
{{}^2}%
$. Theoretical estimates of the critical current density for a synthetic Fe/Cr
antiferromagnet suggest that this effect may be attributed to current-induced
domain-wall motion that occurs via spin transfer torques.

\end{abstract}
\volumeyear{ }
\volumenumber{ }
\issuenumber{ }
\eid{ }
\date{July 14, 2008}
\received[Received text]{date}

\revised[Revised text]{date}

\accepted[Accepted text]{date}

\published[Published text]{date}

\startpage{101}
\endpage{107}
\maketitle

Mutual interaction between electron spin current and ferromagnetic order has
given rise to a variety of spintronic effects and devices. Well known examples
are the giant magnetoresistance (GMR) effect (with ferromagnetic layer moments
acting on electron spin) \cite{Baibichi,Grunberg}, and spin-torque phenomena
\cite{Tsoi,Myers,Slonczewski,Berger} where \textquotedblleft
free\textquotedblleft\ ferromagnetic layer moments are inverted (excited) or
domain walls (DWs) are being displaced
\cite{grollier:509,0295-5075-65-4-526,PhysRevLett.92.077205,klaui:106601,ravelosona:117203}
by a high density spin current flowing through the nanostructure. These
phenomena involve rich new physics and have demonstrated enormous potential
for applications.

The main problem which faces the spin-torque is a need to decrease the
critical currents which induce the effect. Spin transfer in perpendicularly
magnetized media, in exchange-biased spin valves, in ferromagnetic
semiconductors \cite{Jiang,Mangin,Yamanouchi}, or using specific periodic
current pulses \cite{Parkin2006}, partially overcome this difficulty. A
radically different approach to solve the problem proposes to explore DC
current-induced torques in antiferromagnetic (AF) materials
\cite{Nunez,Wei,Urazhdin}. Theoretical estimates suggest the possibility to
substantially reduce the critical currents and influence the domain structure
of antiferromagnetic films already at critical currents above $10^{5}A/cm%
{{}^2}%
$ which are about one order of magnitude below the best reported for the
ferromagnetic structures.

Creation of DWs in antiferromagnets is not energetically favorable. For
synthetic antiferromagnets, such as [Fe/Cr]$_{n}$ magnetic multilayers (MMLs),
the influence of a magnetic field on the DWs is mainly due to the presence of
biquadratic coupling allowing effective displacement of the antiferromagnetic
domain walls by the external fields \cite{Aliev}. A relatively low DW pinning
energy in fully epitaxial Fe/Cr MMLs \cite{Symposium} gives rise to
appreciable domain wall magnetoresistance (DW-MR) only at temperatures below
100K \cite{Europhysics}.

Here we investigate the interaction of DC currents with domain walls in
epitaxial antiferromagnetically coupled [Fe/Cr]$_{10}$ multilayers in a wide
temperature range below 300 K and with current densities up to $5\times
10^{5}A/cm%
{{}^2}%
$, by using simultaneous measurements of domain wall magnetoresistance and low
frequency noise. While the DW-MR remains practically unchanged for the current
densities under investigation, a transition from excess DW induced 1/f noise
at low current densities to a suppressed noise at current densities
\emph{j}$_{c}$\emph{ }exceeding $2\times10^{5}A/cm%
{{}^2}%
$ is observed. This unexpected observation could be attributed to
current-induced domain-wall motion above the critical current density
\emph{j}$_{c}$\emph{ }, which coarsens the DWs and reduces the DW-related
noise. Current-induced DW noise in ferromagnetic Co line structures is found
to remain qualitatively unchanged for densities up to $2\times10^{6}A/cm%
{{}^2}%
$ indicating a weaker influence of current on DWs.

Two different epitaxial [Fe(001)/Cr(001)]$_{10}$ multilayers with the same Cr
thickness of 13\AA \ (providing the maximum antiferromagnetic coupling) and
with different Fe thicknesses of 24\AA \ (MML1) and of 12\AA \ (MML2) were
deposited in a molecular beam epitaxy system on MgO(100) substrates held at
50$%
{{}^\circ}%
$C. Detailed descriptions of the sample preparation and characterization (XRD,
electron transport, etc.) may be found in Refs.~\cite{Europhysics,Schad}. The
samples were patterned to a line structure configuration along the easy (001)
axis with a length of $500$ $\mu m$, a width of $10$ $\mu m$, and a distance
between pairs of oppositely situated voltage probes of $400$ $\mu m$.
Experiments were carried out with the DC current parallel to the external
magnetic field and to the magnetically easy (001) axis. Details of the growth
and characterization of 21 nm thick, $10$ $\mu m$ wide and $700$ $\mu m$ long
polycrystalline Co line structures can be found in Ref.~\cite{Brems}.\ The
setup employed in electron transport and low-frequency noise measurements has
been described previously \cite{APL,R.Guerrero}.

Fig.~1a shows the low-temperature in-plane magnetoresistance\ in samples MML1
and MML2 measured up to high magnetic fields to ensure complete parallel
alignment of the Fe layers. The high GMR values $(60-80\%)$ point to the
strong antiferromagnetic coupling and confirm a good epitaxy. We have found,
however, that the saturation field $H_{s}$ is much better defined for MML1,
indirectly indicating its enhanced crystalline order in comparison with MML2.
In agreement with previous observations, both types of multilayers show an
appreciable DW magnetoresistance only at temperatures below 100K
\cite{Symposium,Europhysics}.

Fig.~1b compares the normalized magnetoresistance measured at a temperature of
77K in MML1 and MML2 for the low magnetic field regions when DWs nucleated at
$H=H_{n}$ and annihilated at $H=H_{a}$ contribute to magnetoresistance. One
clearly observes a less defined DW annihilation field for MML2 in comparison
to MML1, which correlates with a much broader field region needed to suppress
completely the antiferromagnetic state (Figs. 1a,b). This further indicates a
higher local structural order in MML1 in comparison to MML2.

We have found that the DW-MR is independent of the applied current densities
within an order of magnitude below the maximum current densities applied.
Figure 1c demonstrates that even with the maximum current densities used at
T=77K $(j=3.3\times10^{5}A/cm%
{{}^2}%
),$ the current induced self-field or heating are negligible and practically
have no influence on the DW magnetoresistance. To finalize discussion of the
low-field electron transport we also mention that the DW-MR is found to be
practically independent of orientation of the current in respect to the
in-plane magnetic field (see Fig.1d). This clearly indicates absence of
anisotropic magnetoresistance, as expected to occur in the strongly
antiferromagnetically coupled multilayers with reduced magnetization.

The voltage noise power $S_{V}$ due to magnetization fluctuations was studied
in the frequency range between $1$ to $25$ $Hz$. The observed $1/f$ noise was
described as $S_{V}\left(  f,H\right)  =\frac{\alpha\left(  H\right)  V^{2}%
}{f}$, with $\alpha$ the Hooge factor and $V$ the average voltage between
potential contacts \cite{Hardner,Kogan}. Figs. 2 and 3 present our main
findings: clear suppression of the DW related excess low frequency noise
(Hooge factor) in MML1 for sufficiently high current densities. Figs. 2 a,b
show data obtained at T=77K:\ when current densities used exceed
$2\times10^{5}A/cm%
{{}^2}%
$, the excess noise observed for low magnetic fields where DWs are created and
propagate is transformed into suppressed noise in the field interval
$H_{n}<H<H_{a}$. Measurements in similar conditions (77K)\ for MML2 reveal
that the Hooge factor reduced by about an order of magnitude and showed no
qualitative changes up to current densities of $5\times10^{5}A/cm%
{{}^2}%
$. This provides further proof for stronger DW pinning due to higher
structural disorder in MML2.

Decreasing the temperature down to 10K substantially reduces the excess DW
induced noise measured in MML1 with low current densities (Fig.3a). This is in
perfect agreement with suppression of the DW\ related excess imaginary
contribution to magnetic susceptibility due to stronger DW pinning at low
temperatures \cite{Aliev}. Noise measured at 10K\ with low current density
($<10^{5}A/cm^{2}$) weakly increases when DWs are formed. However, for current
densities exceeding $2\times10^{5}A/cm^{2}$ a strong reduction of the
normalized noise for the field region where DWs are created and propagate is
again observed. Figs.3b,c compare the dependence of the Hooge factor on
current density for the field regions with $(H_{n}<H<H_{a})$\ and
without\ $\left(  \text{i.e. for }H>H_{a}\right)  $ the presence of DWs. For
T=77K one observes a crossover from excess to suppressed DW noise while for
T=10K this crossover is shifted to lower current densities. We believe that
some reduction of the current-induced suppression of the DW related noise at
the highest current densities and low temperatures (Fig.3c)\ could be due to
Joule heating and Oersted field effects estimated to be of about 10K and
roughly 5 \emph{Oe} for densities of $4\times10^{5}A/cm^{2}$. We finally
mention that qualitatively\ similar current-induced changes in the normalized
low frequency noise have been observed when experiments have been carried out
with DC current along the easy axis but perpendicular to the external magnetic field.

It was recently observed that the DW\ induced noise\ measured through the
field region where DWs are formed, propagate and annihilate in
quasi-equilibrium conditions, approximately scales with absolute values of the
derivative of resistance \textit{vs.} magnetic field, i.e. $\alpha
(H)\varpropto\frac{\partial R}{\partial H}$ \cite{Nowak}. Figure 3a compares
the field dependence of $\partial R/\partial H$ and Hooge factor measured at
10K\ with a current density of \emph{j=}$2.3\times10^{5}A/cm%
{{}^2}%
$. Clearly, the Hooge factor and $\partial R/\partial H$ show qualitatively
different field dependences close to the region where DWs are formed and
propagate. This observation indicates strongly non-equilibrium DW\ induced
noise at high current densities and also is an indirect proof of setting in
motion the DWs in Fe/Cr multilayers by current. Interestingly, we have found
that the simple relation between $\partial R/\partial H$\ and Hooge factor is
reasonably well fulfilled for the noise measurements in the ferromagnetic Co
line structure in the field region where DWs are nucleated and propagate up to
current densities of about $4\times10^{6}A/cm%
{{}^2}%
$\ (Fig. 4). Qualitatively similar behavior was observed both for 300K and
77K.\ For current densities exceeding $10^{6}$A/cm$%
{{}^2}%
$\ the Oersted field created by the transport current (roughly 12 Oe for the
maximum used current density) and Joule heating (being above 10K for the
maximum DC currents employed)\ may, however, affect both magnetotransport
($\partial R/\partial H$) and low frequency noise in the ferromagnetic Co line
structures. This influence is evidenced in the gradual reduction of the domain
wall depinning field and related displacement of both maxima in $\partial
R/\partial H$ and in Hooge factor above $10^{6}A/cm%
{{}^2}%
$ (see inset to Fig. 4).

We believe our experimental results indicate current-induced motion of the
domains in the magnetic microstructure in the Fe/Cr multilayers, as we now
explain. First of all, the predominant domain walls are those in the Fe layers
induced by local variation of the Cr layer thickness, that causes alternating
exchange coupling between the Fe layers \cite{Schmidt}, as well as extrinsic
pinning for the domain walls. Secondly, because the geometry is
current-in-plane of the layers we assume that within each layer the same
mechanisms prevail that cause current-induced domain-wall motion in
ferromagnetic metallic wires. In the presence of extrinsic pinning the
critical current for depinning a domain wall is then given by
\cite{Tatara,Ghen} $j_{c}\sim\left\vert e\right\vert V_{pin}\lambda/\left(
\hbar\beta A\xi\right)  $, where \emph{e} is the electron charge, $V_{pin}$ is
a typical pinning energy for the domain wall, and \textit{A} is the
cross-sectional area perpendicular to the current direction of a single Fe
layer. For our system parameters we find, taking $V_{pin}/k_{B}$ equal to the
temperature below which DW magnetoresistance is observed ($\sim100$ K), that
$j_{c}\sim\lambda/\beta\xi\times10^{5}$ $A/cm%
{{}^2}%
$ in terms of the width $\lambda$ of the domain wall and the typical range
$\xi$ of the pinning potential. The critical current is further determined by
the dimensionless parameter $\beta$ that characterizes the degree to which
spin is not conserved in the spin transfer process, and/or the degree to which
the spin transfer torque is not adiabatic. In our system we expect that
$\lambda/\xi$ is of the order of $0.1-0.01$ \cite{Schmidt}. Microscopic
calculations \cite{Tserkovnyak,Kohno,Duine,Piechon}\ indicate that $\beta$ is
of the same order as, though generally not precisely equal to, the Gilbert
damping constant $(\sim0.1-0.01)$\ and we conclude that the estimated critical
current is in rough agreement with our experimental findings. From this
theoretical picture we also conclude that the critical current for moving
domain walls is lower in multilayer Fe/Cr than in ferromagnetic wires, mainly
because the pinning energy of the domain walls is smaller due to the
antiferromagnetic structure. To further corroborate the picture of extrinsic
pinning, we remark that estimating the so-called intrinsic critical current
for domain-wall motion \cite{Tatara} from the saturation magnetization of one
Fe layer $(\sim2T)$ yields a critical current that is several orders of
magnitude higher than the current at which the suppression of $1/f$ noise is observed.

Let us finally discuss the possible origin of the unusual minimum in the
normalized $1/f$ noise at high current densities and between the
DW\ nucleation and depinning fields. When current-induced torques overcome the
DW pinning energy and for high current densities the \emph{DWs are set in
motion,} their continuous displacement could essentially reduce a local
variation of the magnetic coupling sign due to 1 ML steps at the Fe/Cr
interface previously reported for Fe/Cr/Fe multilayes \cite{Schmidt}. This
picture explains the substantial suppression of magnetic $1/f$ noise in the
vicinity of the depinning field with current densities $j_{c}>2\times10^{5}$
$A/cm^{2}$ and is supported by the broad maxima in noise at low temperatures
(T=10 K)\ and low currents (Fig. 3c). Indeed, weakly mobile nucleated domains
could enhance magnetic frustration and increase the low frequency noise.

In conclusion, detailed studies of domain wall magnetoresistance and noise in
antiferromagnetically coupled Fe/Cr multilayers reveal an unexpected
dependence of domain wall induced low frequency noise on the current density.
This observation could be attributed to current induced domain wall motion.
Simple estimates confirm the possibility of a strong reduction of the critical
currents needed to move domain walls in synthetic Fe/Cr antiferromagnets in
comparison with ferromagnetic line structures.

\begin{acknowledgments}
It is a great pleasure to acknowledge Allan MacDonald and Gen Tatara for
helpful conversations. This work is supported by Spanish MEC (MAT2006-07196,
MAT 2006-28183-E) and the Stichting voor Fundamenteel Onderzoek der Materie
$(FOM)$ and the Nederlandse Organisatie voor Wetenschaplijk Onderzoek $(NWO).$
\end{acknowledgments}

\bigskip

\section{Figure Captions}

\section{}

Fig 1. (a) Comparison of giant magnetoresistance GMR=$[R(H)-R(60$
$\mathrm{kOe})]/R(60$ $\mathrm{kOe})\times100$ measured at 10K for MML1 and
MML2. The arrows indicate the corresponding saturation fields $H_{s}.$ (b)
Normalized to zero field $\left(  DW-MR\left(  0\right)  =[R(H)-R(0\text{
}\mathrm{Oe})]/R(0\text{ }\mathrm{Oe})\right)  $ value magnetoresistance
measured for MML1 and MML2 at $77K$. (c) Domain wall magnetoresistance
DW-MR=$[R(H)-R(400$ $\mathrm{Oe})]/R(400$ $\mathrm{Oe})\times100$ for MML1 and
MML2 measured at $T=77K$ with different current densities used for the noise
measurements. The arrows indicate the DW nucleation ($H_{n}$ ) and
annihilation ($H_{a}$) fields. (d) Low field magnetoresistance measured at
$T=77K$ with zero bias for MML1 with current either parallel or perpendicular
to the external magnetic field.\bigskip\bigskip

Fig 2.(a) Dependence of the Hooge factor $\alpha$ on the magnetic field
measured for MML1 at 77K with different applied current densities. (b) Hooge
factor vs. magnetic field for MML1 at 77K with two different applied current
densities shown in logarithmic scale. The normalized noise vs. field
measurements obtained with a current density of $j=3.3\times10^{5}A/cm%
{{}^2}%
$ are shown both for increasing and decreasing magnetic field. The experiments
have been carried out with DC current parallel to the external magnetic field
and to the magnetic easy (001) axis. \bigskip

Fig 3 (a)\ Dependence of the Hooge factor $\alpha$ (in logarithmic scale) on
the magnetic field for MML1 measured at T=10K with two different applied
current densities. For comparison, solid line shows absolute values of
derivative $\partial R/\partial H$measured with current density of
$2.3\times10^{5}A/cm%
{{}^2}%
$. The vertical arrows mark the regions of nucleation\ and annihilation\ of
the domain walls determined from low-field magnetoresistance.
(b,c)\ Dependence of the Hooge factor in the current density measured in MML1
with and without domain walls for two different magnetic fields at
temperatures of 77K\ (part b)\ and of 10K\ (part c).

Fig 4 Hooge factor $\alpha$ vs. magnetic field measured in the ferromagnetic
Co line structure at 77K with current density of $j=3.3\times10^{5}A/cm%
{{}^2}%
$. The insert shows the field values of the maximum in Hooge and related
maxima of the derivative $\partial R/\partial H$ (as indicated by dashed
circle) as a function of current density.%

\raisebox{-0cm}{\includegraphics[
height=7.2159cm,
width=8.7052cm
]%
{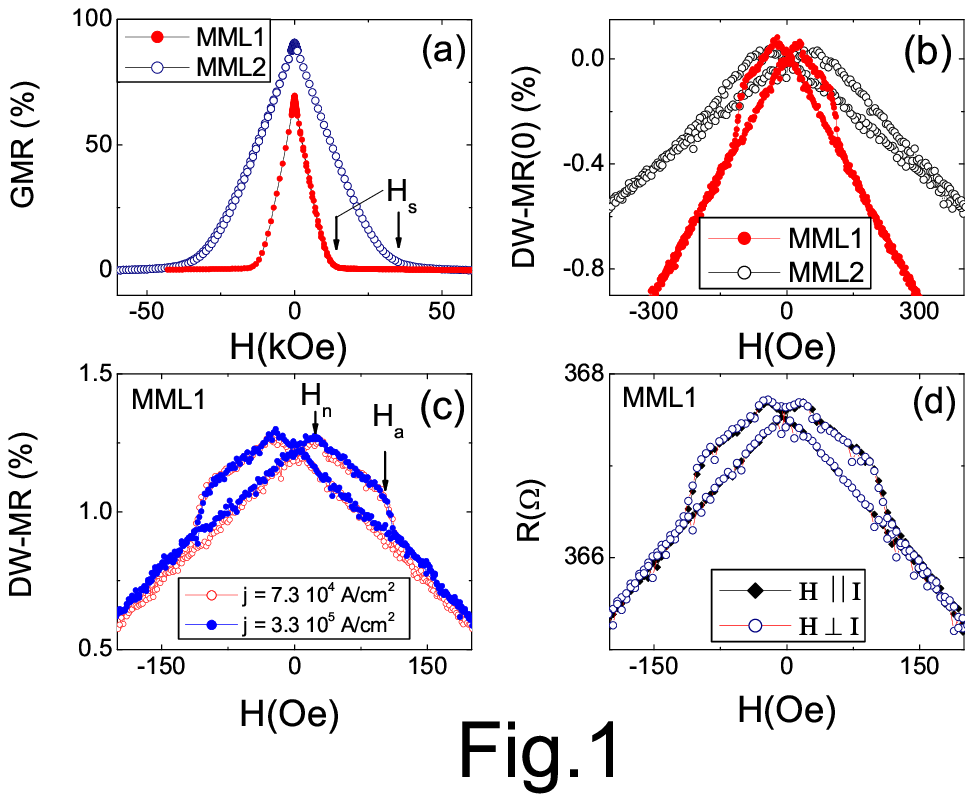}%
}%
%

\raisebox{-0cm}{\includegraphics[
height=4.4416cm,
width=8.4768cm
]%
{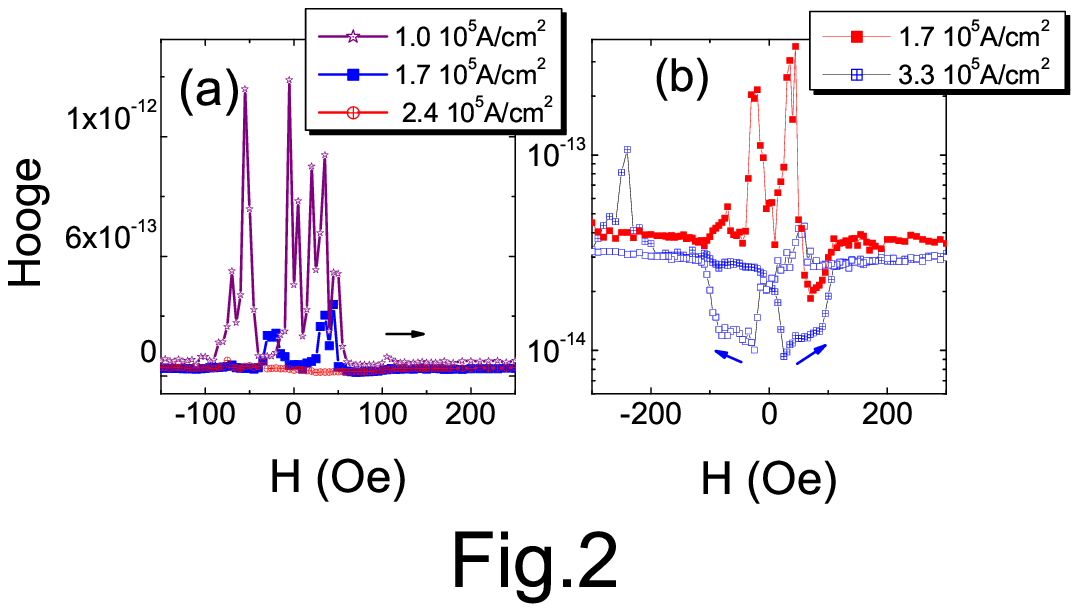}%
}%
%

\raisebox{-0cm}{\includegraphics[
height=5.8276cm,
width=8.5536cm
]%
{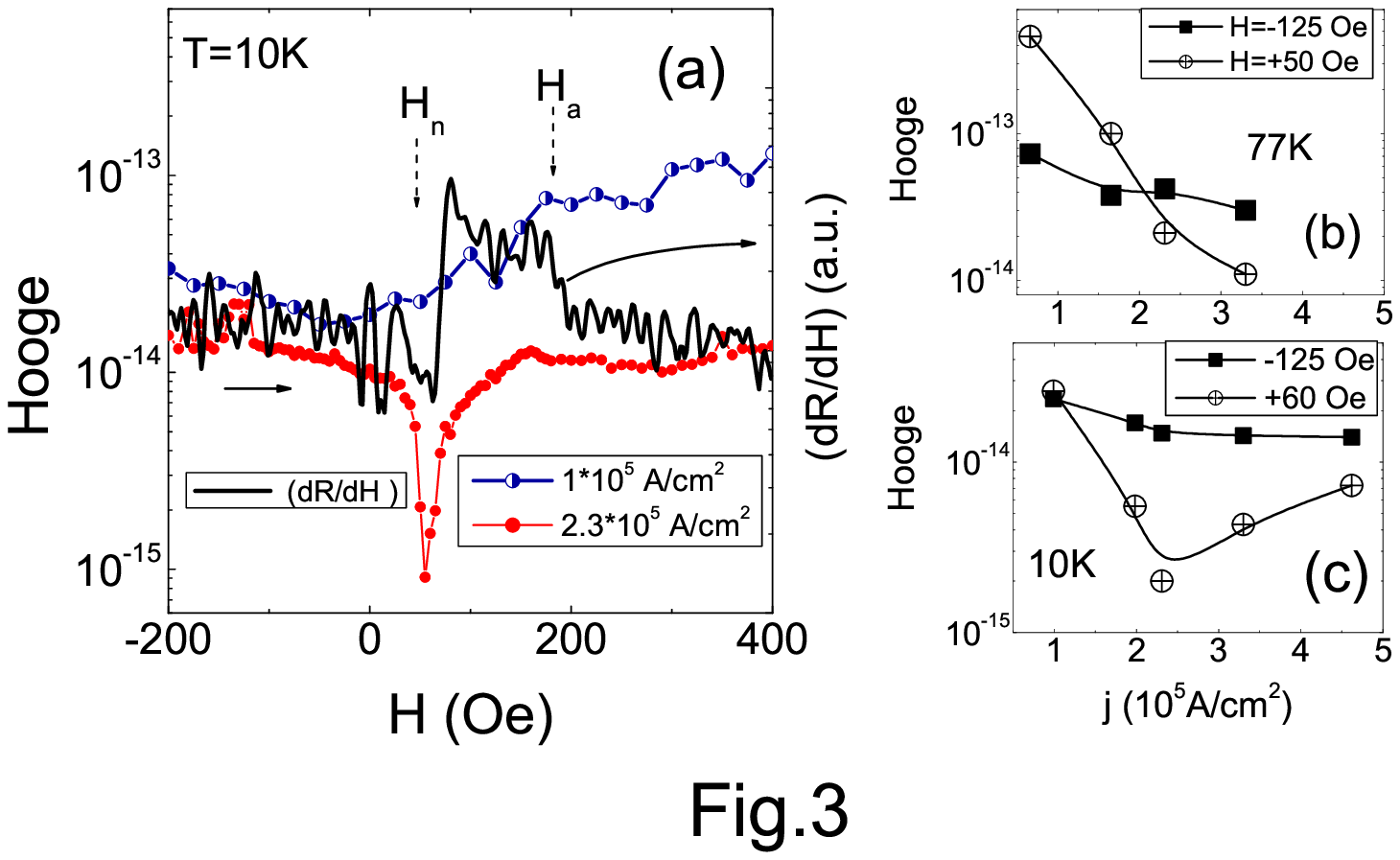}%
}%
%

\raisebox{-0cm}{\includegraphics[
height=7.262cm,
width=8.613cm
]%
{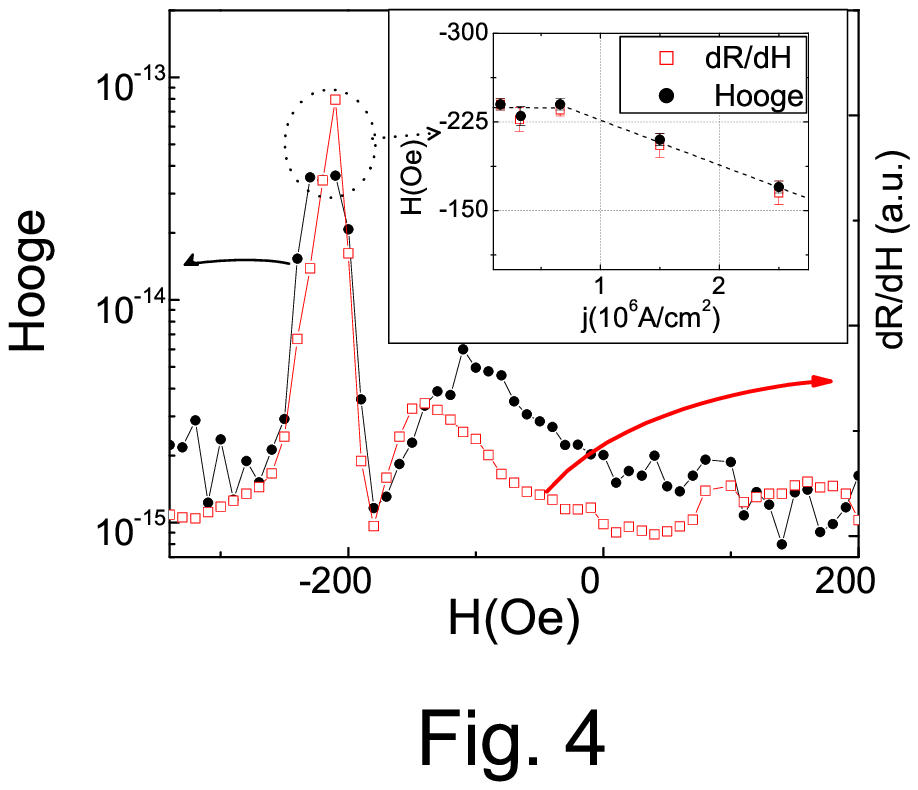}%
}%

\end{document}